# Thickness-dependent structural, magnetic and transport properties of epitaxial Co$_2$FeAl Heusler alloy thin films

Wenhong Wang, Enke Liu, Yin Du, Jinglan Chen, and Guangheng Wu

*Beijing National Laboratory for Condensed Matter Physics, Institute of Physics, Chinese Academy of Sciences, Beijing 100190, P. R. China*

Hiroaki Sukegawa, Seiji Mitain, and Koichiro Inomata

*National Institute for Materials Science (NIMS), 1-2-1 Sengen, Tsukuba, 305-0047, Japan*

**Abstract:** We report on a systematic study of the structural, magnetic properties and the anomalous Hall effect, in the Heusler alloy Co$_2$FeAl (CFA) epitaxial films on MgO(001), as a function of film thickness. It was found that the epitaxial CFA films show a highly ordered *B*2 structure with an in-plane uniaxial magnetic anisotropy. An analysis of the electrical transport properties reveals that the lattice and magnon scattering contributions to the longitudinal resistivity. Independent on the thickness of films, the anomalous Hall resistivity of CFA films is found to be dominated by skew scattering only. Moreover, the anomalous Hall resistivity shows weakly temperature dependent behavior, and its absolute value increases as the thickness decreases. We attribute this temperature insensitivity in the anomalous Hall resistivity to the weak temperature dependent of tunneling spin-polarization in the CFA films, while the thickness dependence behavior is likely due to the increasing significance of interface or free surface electronic states.

-------------------------------------------------

1. Introduction

The performance of spintronics devices depends on spin polarization of the current. Therefore, highly spin polarized current source is strongly desired in spintronics devices. The use of half-metallic ferromagnets (HMFs) is a typical approach for this purpose. HMFs are characterized by a metallic density of states at Fermi level ($E_F$) for one spin channel, while the states for the other spin channel display a gap at $E_F$, leading to 100% spin polarization.[1] Among some kinds of HMFs candidates, Co-based full-Heusler alloys with a chemical form of $Co_2YZ$ and $L2_1$ structure have been growingly investigated since a large tunneling magnetoresistance (TMR) observation at RT using a $Co_2Fe_{0.4}Cr_{0.6}Al$ (CCFA) electrode in magnetic tunnel junctions (MTJs).[2] So far, more than hundred percent TMR has been reported in MTJs using half-metallic Co-based full-Heusler alloy electrodes such as $Co_2MnGe$ (CMG),[3] $Co_2FeAl_{0.5}Si_{0.5}$ (CFAS),[4-6] and $Co_2MnSi$ (CMS).[7-9] In general, however, TMR in MTJs using the full-Heusler alloy electrodes largely decreases with increasing temperature, especially for CMS electrode, of which has been attributed to the $E_F$ near the bottom edge of the conduction band of CMS,[10] nonquasiparticle (NQP) states appearing just above $E_F$ and the creation of interface states at $E_F$ within the half-metallic gap.[11] As a result, the spin-flipping process through the interface states mediated by magnon excitation.[12]

Very recently, we have demonstrated a high tunneling spin polarization in MTJs using a $Co_2FeAl$ (CFA) electrode, achieving a high TMR ratio of 330% at room temperature due to coherent tunneling.[13,14] We found that the temperature dependence of TMR in epitaxial MTJs using the full-Heusler alloy electrodes can be explained quantitatively using Julliere's model[15] for TMR (TMR ratio = $2P_1P_2 / (1 - P_1P_2)$) and the spin wave excitation model for the spin polarization $P(T) = P_0 (1-\alpha T^{3/2})$,[16,17] where $P_0$ is the spin polarization at 0 K and $\alpha$ is a constant dependent on magnetic materials for electrodes and the interface structure. The analysis exhibits that the CFA film has an almost 100% spin polarization at 0 K, showing a very weak temperature dependent behavior. However, the magnetic tunnelling junction experiments will likely introduce additional, barrier dependent and/or interface contributions to the TMR which will definitely influence intrinsic spin polarization of CFA electrodes. In order to clarify such effects from possible intrinsic contributions to the temperature dependence of the tunneling currents, the investigations of the magnetic and the transport properties is indispensable. Moreover, from the application point of view, the integration of CFA as a ferromagnetic electrode in spintronic devices also requires a precise knowledge and control of its transport properties. For the sputtered Heusler alloy films, the degree of ordering can be controlled through different annealing process, and therefore most previous studies have been devoted into the electronic magnetotransport properties for the sample states with different postannealing temperatures.[18-20] Nevertheless, thickness-dependent degree of ordering and the transport properties of films have been suggested to play an important role in the transport spin-polarization of Heusler alloys. In this Brief Report, we will study the effect of the variation of film thickness on the structural, magnetic properties, and the anomalous Hall effect (AHE) in the high quality sputtered CFA Heusler thin films.

## 2. Experimental procedures

Thin films of $Co_2FeAl$ (CFA), with thickness of 5, 10, 20, 30, 50, and 80 nm, were grown on MgO(001) single crystal substrates using ultra-high vacuum magnetron sputtering system with the base pressure of below $8\times10^{-8}$ Pa. Before the deposition, the MgO (001) substrates were heated to 700 °C for 1 h in-situ under vacuum. After cooling down to RT (23 °C) under vacuum, a 20-nm-thick MgO buffer layer was deposited on the substrate. The MgO buffer layer was deposited by rf sputtering directly from a sintered MgO target under an Ar pressure of 10 mTorr. After depositing the buffer layer, CFA films were deposited subsequently from a stoichiometric Co-Fe-Al (Co: 50.0%, Fe: 25.0%, Al: 25.0%) target. The Ar pressure during sputtering was 1.0 mTorr and typical deposition rate was $2\times10^{-2}$ nm/s for CFA. We found that the as-deposited CFA film composition was $Co_{2.20}Fe_{1.00}Al_{0.91}$ through inductively coupled plasma analysis (ICP). To achieve a highly B2-ordered structure, an *in-situ* post-deposition annealing at 480 $^0$C was carried out after the CFA films deposition. Finally, the films are capped with 2 nm of MgO to prevent the oxidation.

The structural properties of the sample were investigated by *ex-situ* X-ray diffraction (XRD) with Cu $K_\alpha$ radiation, where out-of-plane ($2\theta/\theta$ scan) and in-plane ($2\theta\chi/\phi$ scan) diffraction measurements were performed to detect diffraction signals from lattice planes of perpendicular and parallel to the sample surfaces, respectively. Magnetic properties were characterized by vibrating sample magnetometer (VSM). Surface morphology and surface roughness were investigated using atomic force microscopy (AFM). The magnetotransport measurements were performed at various temperatures between 2K and 300K using a physical property measurement system (PPMS). For avoiding magnetoresistive contributions due to misalignment of the contracts, the measurements were accomplished with positive and negative magnetic field.

## 3. Results and Discussion

Figure 1(a) shows X-ray $\theta$-$2\theta$ diffraction patterns of the series of 480 °C annealed CFA films prepared on MgO-buffered MgO (001) substrates. In addition to peaks from MgO (001) substrate, one can see that all the fabricated CFA films exhibit only the (002) and (004) peaks, indicating perfect (001)-oriented growth on the MgO (001) surface. All the films are determined to be the *B*2-ordered structure due to the presence of (002) peaks. It is well known that the ordered structure of Heusler alloy films is characterized whether there exists either a diffraction peak from the (111) superlattice reflection corresponding to the *L*2$_1$–ordered structure or a peak from (222) superlattice reflection corresponding to *B*2–ordered structure. We therefore further carried out the X-ray in-plane $\phi$-scan measurements to confirm the degree of order for the prepared CFA films. As a result, in Figure 1 (b), we show the X-ray pole figures of (222) for the series of CFA films annealed at 480 $^o$C. The distinct *B*2–ordered (222) superlattice peaks with fourfold symmetry with respect to $\phi$ were clearly observed for the CFA films. This provides direct evidence that these CFA films grow epitaxially and have the *B*2–ordered structure, which implies existence of disorder in the atomic site between Fe and Al. In addition, the peak intensity of (222) increases with increasing $T_a$, which indicates the crystal structure of the CFA film has been improved, and the level of the improvement strongly depends on the thickness. Consequently, the epitaxial relationship, as shown in Fig. 1(c), is found to be CFA (001)[100]//MgO (001)[110]. This relationship is reasonable in view of the relatively small lattice mismatch (-2.9%) between CFA (001) and MgO (001) on 45° in-plane rotation. The in-plane and out-of-plane lattice constants of the epitaxial CFA films were determined using CFA (004) and CFA (404) peaks, respectively. As shown in Fig. 1 (d), the lattice constants of both the in-plane and out-of-plane show a very weak thickness dependence behavior and both values approached approximately to the bulk value (0.573 nm) reported experimentally.[21]

In order to fabricate high-quality magnetic tunneling junctions devices, a lower ferromagnetic electrode with small surface roughness must be prepared. We further characterized the surface roughness for the CFA films using AFM. Figures 2(a)-(e) show the AFM images of CFA films with various thicknesses. One can see that the 5-nm-thicknes CFA film shows a very small grain size; however, for the CFA film thicker than 30 nm, the grain sizes become very large with random orientation, which leads to a relative rough surface. In Fig. 2 (d), we show the average surface roughness (*Ra*) and peak-to-valley (*P-V*) values for the series of CFA as a function of the thickness. Surprisingly, we found that both *Ra* and *P-V* values are almost thickness independent for the CFA films thinner than 50 nm. As a result, a remarkable flat surface with both *Ra* of 0.11 nm and *P-V* of 1.10 nm was achieved by 480 °C post-annealing. We should point out that the surface roughness is comparable to the CFA films prepared on Cr-buffered MgO (001) substrates. So in this sense, we could conclude that the highly *B*2 ordering and the flat surface after post-deposition annealing indicate the use of the MgO buffer for CFA layer is a promising approach for the application of spin injection source.

The magnetic properties of the epitaxial CFA films were investigated by VSM at RT. Figure 3 shows magnetization vs. applied magnetic field for magnetic fields applied in-plane along the [100] and [110] crystallographic orientations of the 30-nm-thickness CFA film. A distinct in-plane magnetic anisotropy is observed in all CFA thin films. The magnetic easy-axis is along the CFA [110] crystallographic direction, due to essentially the full magnetic moments at remanence with sharp and square hysteresis loops. The hard-axis is along [100] direction since rounded hysteresis loops are observed, and the magnetic moments at remanence is clearly reduced relative to their saturation values.

It is worthy to note that, although an in-plane uniaxial anisotropy has been observed in several Heusler alloy films gown on semiconductors of the zinc-blende structure,[22-25] however its microscopic origin remains an open question. There have been many theoretical efforts to explain the origin of the in-plane uniaxial anisotropy. For example, it might be due to a reconstruction of the semiconducting substrates, formation of an interface alloy, or the most likely anisotropic interfacial bonds. In the case of our CFA/MgO (001) films, however, the uniaxial anisotropy is independent of the thickness of CFA films. As a result, we found that the uniaxial behavior of the in-plane magnetic anisotropy persists up to 80 nm, and therefore it can not be related to any shape anisotropy of the incipient growing film. In a recent publication, Gabor *et. al.*,[20] have reported that, for the epitaxial CFA films on MgO (001) substrates, the in-plane uniaxial anisotropy can be decomposed into the two uniaxial in-plane anisotropies with one easy axis of magnetization along the [110] and the other along [100] directions, respectively. However, we did not observe the four-fold in-plane uniaxial anisotropy in our CFA films. We therefore suggest that this thickness-independent in-plane uniaxial anisotropy has a magnetostatic origin related to lattice misfit between MgO and CFA films. To verify this assumption, we have also fabricated films using a 20-nm thick Cr-buffer layer. In these samples the substrate morphology influence is expected to vanish. As a result, we did not observe any in-plane magnetic anisotropy in the CFA films prepared on Cr-buffered MgO (001) substrates, which may be due to the small lattice misfit between Cr and $Co_2FeAl$ (1.9%). This result comes to further support that the uniaxial magnetic anisotropy is a result of an anisotropic interfacial bonds induced by the lattice misfit between MgO and CFA films. Nevertheless, further explanation would require identifying some symmetry-breaking mechanism by using the miscut of the substrate.

Figure 3 (c) shows the saturation magnetic moments $M_s$ and the coercivity $H_c$ as a function of film thickness of $t_{CFA}$. All the data were derived from the magnetic hysteresis loops measured by VSM at RT with a magnetic field applied in-plane along the easy-axis, i.e., [110] direction of CFA films [see Fig. 3 (b)]. We found that the $M_s$ first increases and then almost keeps unchanged at $t_{CFA}>10$ nm, while the $H_c$ first increases slowly and then increases rapidly at $t_{CFA}> 50$ nm. The thickness dependence behavior of $M_s$ is similar to the Co$_2$MnSi Heusler films grown on GaAs (001) substrates, [25, 26] which indicate the degree of ordering increases with increasing $t_{CFA}$. The rapid increase of $H_c$ at $t_{CFA}> 50$ nm can be explained due to the formation of large random oriented CFA grains as shown in the above AFM measurements.

The temperature dependence of the normalized longitudinal resistivities ($\rho_{xx}$) of all the CFA films up to room temperature is shown in Fig. 4 (a). As expected, the resistivity decreases with decreasing temperature in the temperature range between 300 and 50 K. However, at temperature below 50 K, as shown in Fig. 4 (b), the resistivity is almost temperature-independent. Here we should point out that in many Heusler alloys films, especially including Mn atoms, the resistivity always shows a local minimum at low temperatures with the resistivity increasing slightly toward lower temperature. [27, 28] However, this type of low-temperature anomaly is not observed in our *B*2-ordered CFA films. The plateau behavior at temperatures below 50 K has also been found in high quality Co$_2$FeSi single crystal, which is attributed to the absence of a one magnon channel and thus half-metallic ferromagnetism. [29] However, previously reported results suggest that CFA is not a true half metal thus further experimental and theoretical study in the understanding of the plateau behavior at low temperature is needed.

On the other hand, a reliable indicator of film purity is the residual resistivity ratio (RRR), i.e., the ratio of the resistivities at 300K and 5 K. As shown in Fig. 4 (c), we plot the calculated RRR as a function of the thickness. We found the values of RRR in the range of 1.1-1.2, which are smaller than the values of the sputtered Heusler films prepared at high substrate temperatures.[27, 30] Moreover, the RRR increase with decreasing thickness of films, indicating that defect scattering is even more dominant in thinner films.

The Hall resistivity ($\rho_{xy}$) in ferromagnetic metals can be commonly expressed as $\rho_{xy} = R_O B + R_S \mu_0 M$, where $R_O$ and $R_S$ are the ordinary and anomalous Hall coefficients, respectively; $B$ is the magnetic field; and $M$ is the magnetization. The first term ($R_O B$) denotes the ordinary Hall effect (OHE),[31] and the second term ($R_S \mu_0 M$) denotes the anomalous Hall effect (AHE).[32] The AHE is proportional to $M$ and conventionally originates from asymmetric scattering process involving a spin-orbit interaction. The effect of the OHE usually appears in the higher fields, while the measured Hall resistivity is dominated by of the AHE in the lower fields.

In figure 5 (a) and (b), we show our results of Hall resistivity, $\rho_{xy}$, as a function of magnetic field $H$ ($H \perp$ thin film) at various temperatures for the two typical thickness of 5 nm and 30 nm, respectively. We found that the $\rho_{xy}$ curves show a clearly nonlinear dependence on the magnetic field, thus indicating that the measured Hall resistivity is dominated by of the AHE in our samples. Both the magnitude of the Hall signal and the temperature dependence are very similar between the two samples. At low $H$, $\rho_{xy}$ increases linearly with the magnetic field due to the linear variation of the magnetization; the slight rounding of $\rho_{xy}(H)$ near saturation filed may be attributed to residual demagnetization in–plane fields extending from imperfections alignment of magnetic moments. This is achieved slightly at high $H$, where the field dependence of the negative ordinary Hall effect survives.

Since $\rho_{xy}$ is dominated by of the AHE in low fields and therefore the OHE contribution to the total Hall effect can be generally neglected. As a result we may reliably identify the AHE at each temperature from the intercept of $\rho_{xy}$ curve to zero field by assuming OHE to be nearly temperature independent. Figure 6 (a) shows the AHE resistivity ($\rho_{AHE}$) of all the CFA films at H=0 as a function of temperature. It was found that the AHE resistivity is weakly temperature dependent, and its magnitude increases with decreasing thickness. These features are consistent with the previous report on the sputtered $Co_2MnSi$ films, where a temperature independent scaling of AHE resistivity was also observed.[26] We attribute the thickness dependence behavior to the increasing significance of interface or surface electronic states. On the other hand, the weak temperature dependent of AHE resistivity may be attributed to the temperature insensitivity of the tunneling spin-polarization for the CFA films as revealed by the previous magnetic tunneling junctions experiments.[17] Recently, by using x-ray magnetic circular dichroism experiments, Klaer *et al.*[33] have investigated the temperature dependence of the spin-resolved unoccupied density of states in the CFA films, and they did not observed a significant difference for surface and bulk related spectra, indicating a similar temperature dependent electronic property of interface and bulk of the CFA films.

In the following we discuss the origin of AHE. It is well known that in the classical model of AHE for ferromagnetic metals, the AHE resistivity is given by $\rho_{AHE} = a\rho_{xx} + b\rho_{xx}^2$,[34] where $\rho_{xx}$ is the longitudinal resistivity and $a$ are $b$ are constants. The linear term in $\rho_{xx}$ is attributed to asymmetric skew scattering of charge carriers, a process which derives from the classical Boltzmann equation.[34, 35] On the other hand, the quadratic term in $\rho_{xx}$ is attributed to asymmetric side jumps which is a purely quantum scattering process.[36] In Fig. 6 (b), we plot $\rho_{AHE}$ as a function of the longitudinal resistivity $\rho_{xy}$ at different temperatures. A clear linear relationship is found for all the CFA films. This implies that the temperature dependence of the AHE in the CFA films is mainly governed by skew scattering. Nevertheless, our experimental results can not indicate whether the AHE is intrinsic or extrinsic as has been pointed out by Lee *et al.*[37] in the ferromagnetic $CuCr_2Se_{4-x}Br_x$ spinel system. Further progress in the understanding of the AHE from Berry curvatures from first principles in Heusler alloys is needed.[38]

## 3. Conclusion

We have studied the structural, magnetic and magnetotransport properties of $Co_2FeAl$ films epitaxial grown on MgO-buffered MgO(001) substrates. We found that single-crystal CFA thin films with high degree of *B*2 ordering and sufficiently flat surface could be obtained on MgO-buffered MgO (001) substrates through magnetron sputtering after post-deposition annealing. The room-temperature saturation magnetization and the coercive field change markedly with the thickness of films. An analysis of the electrical transport properties reveals that the lattice and magnon scattering contributions to the longitudinal resistivity and a plateau behavior at temperatures below 50 K has also been found in all CFA films. The scaling behavior of anomalous Hall effect independent on the thickness of films, which is found to be dominated by skew scattering only. Moreover, the AHE resistivity shows weakly temperature dependent behavior in all the CFA films, and its absolute value increases as the thickness decreases. We attribute this temperature insensitivity in the AHE resistivity to the weak temperature dependent of tunneling spin-polarization in the CFA films, while the thickness dependence behavior is likely due to the increasing significance of interface or free surface electronic states. The possibility to grow high quality CFA expitally on MgO(001) over a wide range of thickness of films and a detailed knowledge of the related physical properties are indispensable for achieving a higher tunnel magnetoresistance ratio, and thus for spintronics device applications.

**Acknowledgement**

This work was supported by National Natural Science Foundation of China Grant Nos. 51071172 and 51021061, and the National Basic Research Program of China (973 Program 2012CB619405).

# References

[1] R. A. Groot, F. M. Mueller, P. G. van Engen, and K. H. J. Buschow, Phys. Rev. Lett. **50**, 2024 (1983).

[2] Ishida, S. Fujii, S. Kashiwagi, and S. Asano, J. Phys. Soc. Jpn. **64**, 2152 (1995).

[3] I. Galanakis, P. H. Dederichs, and N. Papanikolaou, Phys. Rev. B **66**, 174429 (2002).

[4] S. Picozzi, A. Continenza, and A. J. Freeman, Phys. Rev. B **66**, 094421 (2002).

[5] K. Inomata, S. Okamura, R. Goto, and N. Tezuka, Jpn. J. Appl. Phys. **42**, L419 (2003).

[6] Y. Sakuraba, M. Hattori, M. Oogane, Y. Ando, H. Kato, A. Sakuma, T. Miyazaki, and H. Kubota, Appl. Phys. Lett. **88**, 022503 (2006).

[7] T. Ishikawa, T. Marukame, H. Kijima, K.-I. Matsuda, T. Uemura, M. Arita, and M. Yamamoto, Appl. Phys. Lett. **89**, 192505 (2006).

[8] T. Marukame, T. Ishikawa, S. Hakamata, K. Matsuda, T. Uemura, and M. Yamamoto Appl. Phys. Lett. **90**, 012508 (2007).

[9] N. Tezuka, N. Ikeda, A. Miyazaki, S. Sugimoto, M. Kikuchi, and K. Inomata, Appl. Phys. Lett. **89**, 112514 (2006).

[10] G. H. Fecher and C. Felser, J. Phys. D: Appl. Phys. **40**, 1582 (2007).

[11] L. Chioncel, Y. Sakuraba, E. Arrigoni, M. L. Katsnelson M. Oogane, Y. Ando, T. Miyazaki, E. Burzo and A. I. Lichtenstein, Phys. Rev. Lett. **100**, 086402 (2008).

[12] P. Mavropoulos, M. Lezaic and S. Blugel, Phys. Rev. B **72**, 174428 (2005).

[13] W. H. Wang, H. Sukegawa, R. Shan, S. Mitani and K. Inomata, Appl. Phys. Lett. **95**, 182502 (2009).

[14] W. H. Wang, E. K. Liu, M. Kodzuka, H. Sukegawa, M. Wojcik, E. Jedryka, G. H. Wu, K. Inomata, S. Mitani, and K. Hono, Phys. Rev. B **81**, 140402(R) (2010).

[15] M. Julliere, Phys. Lett. A **54**, 225 (1975).

[16] R. Shan, H. Sukegawa, W. H. Wang, M. Kodzuka, W. F. Li, T. Furubayashi, S. Mitani, K. Inomata, and K. Hono, Phys. Rev. Lett. **102**, 246601 (2009).

[17] W. H. Wang, H. Sukegawa, and K. Inomata, Phys. Rev. B 8**2**, 092402 (2010).

[18] H. Schneider, E. Vilanova, B. Balke, C. Felser, and G. Jakob, J. Phys. D: Appl. Phys. 41, 084012 (2009).

[19] I. M. Imort, P. Thomas, G. Resiss, and A. Thomas, J. Applied Phys. 111, 07D313 (2012).

[20] M. S. Gabor, T. Petrisor Jr., C. Tiusan, M. Hehn, and T. Petrisor, Phys. Rev. B 84, 134413 (2011).

[21] T. M. Nakatani, A. Rajanikanth, Z. Gercsi, Y. K. Takahashi, K. Inomata, and K. Hono, J. Appl. Phys. **102**, 033916 (2007).

[22] T. Ambrose, J. J. Krebs, and G. A. Prinz, Apply. Phys. Lett. **76**, 3280 (2000).

[23] J. W. Dong, L. C. Chen, C. J. Palmstrøm, R. D. James, and S. McKernan, Appl. Phys. Lett. **77**, 4190 (2000).

[24] J. Lu, J. W. Dong, J. Q. Xie, S. McKernan, C. J. Palmstrøm, and Y. Xin, Appl. Phys. Lett. **83**, 2393 (2003).

[25] W. H. Wang, M. Przybylski, W. Kuch, L. I. Chelaru, J. Wang, Y. F. Lu, J. Barthel, H. L. Meyerheim, and J. Kirschner, Phys. Rev. B **71**, 144416 (2005).

[26] W. R. Branford, L. J. Singh, Z. H. Barber, A. Kohn, A. K. Petford-Long, W. Van Roy, F. Magnus, K. Morrison, S. K. Clowes, Y. V. Bugoslavsky, and L. F. Cohen, New Journal of Physics. 9, 42 (2007).

[27] M. Obaida, K. Westerholt, and H. Zabel . Phys. Rev. B **84**, 184416 (2010).

[28] W. R. Branford, S. K. Clowes, Y. V. Bugoslavsky, S. Gardelis, J. Androulakis, J. Giapintzakis, C. E. A. Grigorescu, S. A Manea, R. S. Freitas, S. B. Roy, L. F. Cohen, Phys. Rev. B **69**, 201305(R) (2004).

[29] C. G. F. Blum, C. A. Jenkins, J. Barth, C. Felser, S. Wurmehl, G. Friemel, C. Hess, G. Behr, B. Buchner, A. Reller, S. Riegg, S. G. Ebbinghaus, T. Ellis, P. J. Jacobs, J. T. Kohlhepp, and H. J. M. Swagten , Appl. Phys. Lett. **95** 161903 (2009).

[30] L. J. Singha. Z. H. Barber, Y. Miyoshi, Y. Bugoslavsky, W. R. Branford, and L. F. Cohen, Appl. Phys. Lett. **84** 2376 (2004).

[31] E. Hall, Am. J. Math. 2, 287 (1879).

[32] E. Hall, "On the new action of magnetism on a permanent electric current," Ph.D. thesis, (Johns Hopkins University, 1880).

[33] P. Klaer, E. A. Jorge, M. Jourdan, W. H. Wang, H. Sukegawa, K. Inomata, and H. J. Elmers Phys. Rev. B **82**, 024418 (2010).

[34] J. M. Luttinger and W. Kohn, Phys. Rev. **95**, 1154 (1954).

[35] J. Smit, Physica 24, 39 (1958).

[36] L. Berger, Phys. Rev. B **2**, 4559 (1970).

[37] W.-L. Lee, S. Watauchi, V. L. Miller, R. J. Cava, and N. P. Ong, Science **303**, 1647 (2004).

[38] J. Kubler and C. Felser, Phys. Rev. B **85**, 012405 (2012).

Figure captions:

FIG. 1. (Color online) XRD for (a) out-of-plane and (b) $\phi$ scans for (222) plan for the series of CFA films prepared on MgO-buffered MgO (001) substrates; (c) Top view of CFA crystal structure. MgO cubic structure is superimposed; (d) In-plane and out-of-plane lattice constant as a function of film thickness.

FIG. 2. (Color online) (a)-(e) show the show the AFM images of CFA films with various thicknesses. (f) Average surface roughness [$Ra$: ■] and peak-to-valley [$P$-$V$: ●] values for the CFAS films as a function of film thickness.

FIG. 3. (Color online) (a) Magnetic hysteresis loops measured at room temperature for 30-nm-thickness CFA film prepared on MgO-buffered MgO (001) substrates. Measurements were carried out at RT with the magnetic field applied in the film plane along [110] and [100] CFA, respectively. (b) Magnetic hysteresis loops measured at room temperature for the various thickness CFA films along [110] CFA. (d) Saturation magnetic moments ($M_s$, solid lines) and coercivity ($H_c$, doted lines) for the CFA films as a function of film thickness $t_{CFA}$.

FIG. 4. (Color online) (a) Temperature dependence of the longitudinal resistivities $\rho_{xx}$ of the series of CFA films; (b) Resistivity as a function of temperature on a logarithmic scale to show the plateau at temperatures below 50 K; (c) The residual resistivity ratio RRR=$\rho_{300K}/\rho_{2K}$ as a function of film thickness $t_{CFA}$.

FIG. 5. (Color online) Isothermal Hall-resisitivity $\rho_{xy}$ of (a) 5nm and (b) 30 nm CFA films, respectively, in magnetic fields applied perpendicular to the film plane. The determination of the anomalous Hall resistivity $\rho_{AHE}$ (T), and the ordinary hall resistivity at $H=M_s$, $\rho_{OH}$ (T) is indicated.

FIG. 6. (Color online) Anomalous Hall effect resistivity ($\rho_{AHE}$) as a function of (a) temperature and (b) longitudinal resistivities $\rho_{xx}$ for the CFA films with various thickness. The solid lines indicate linear fits.

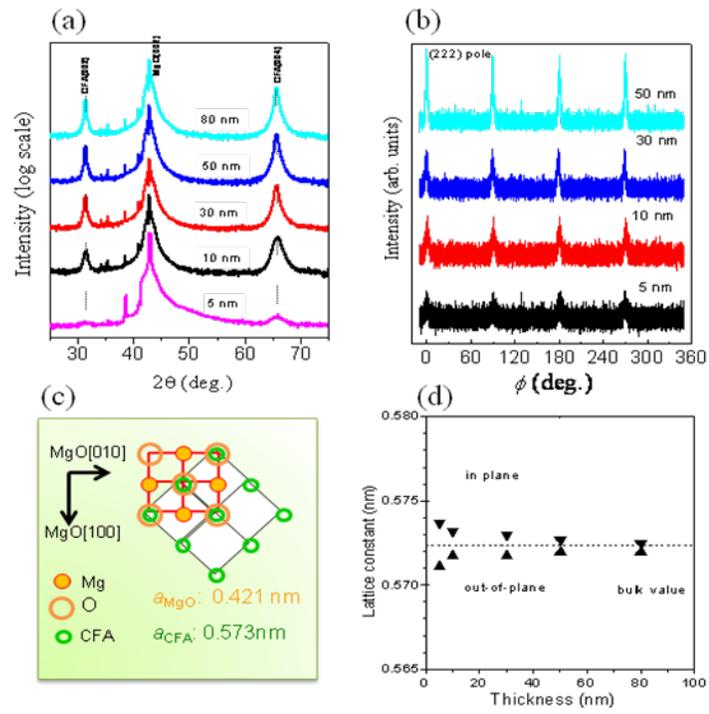

Figure 1

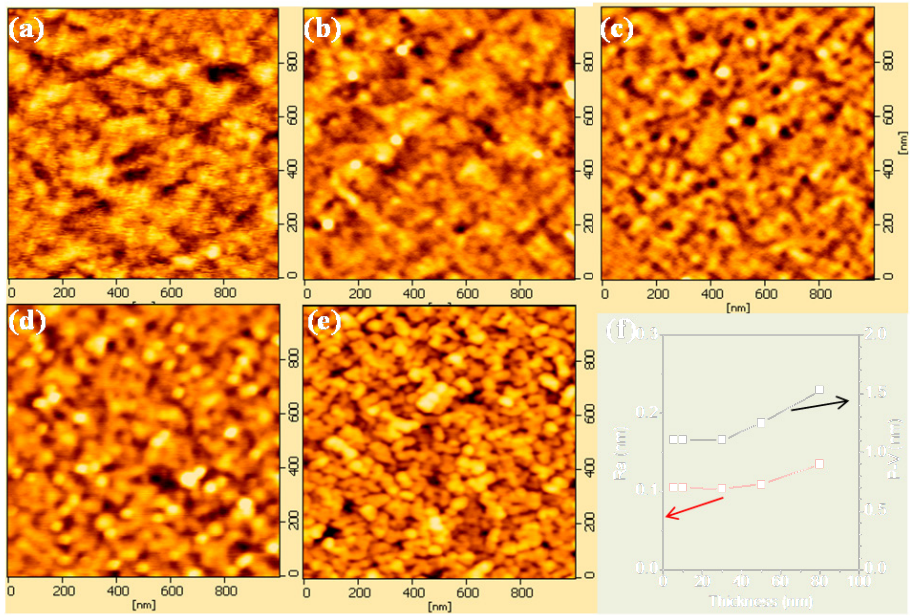

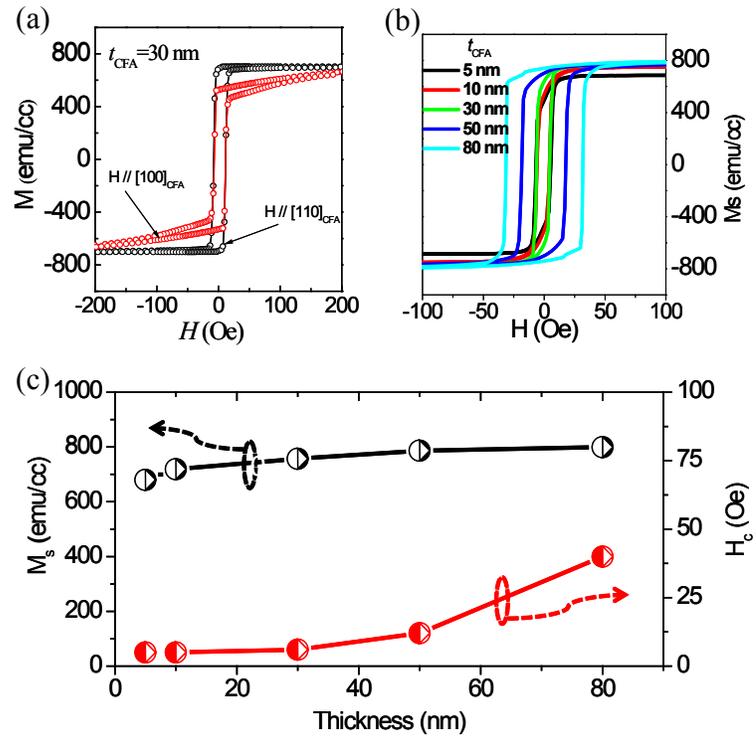

FIG. 3
20

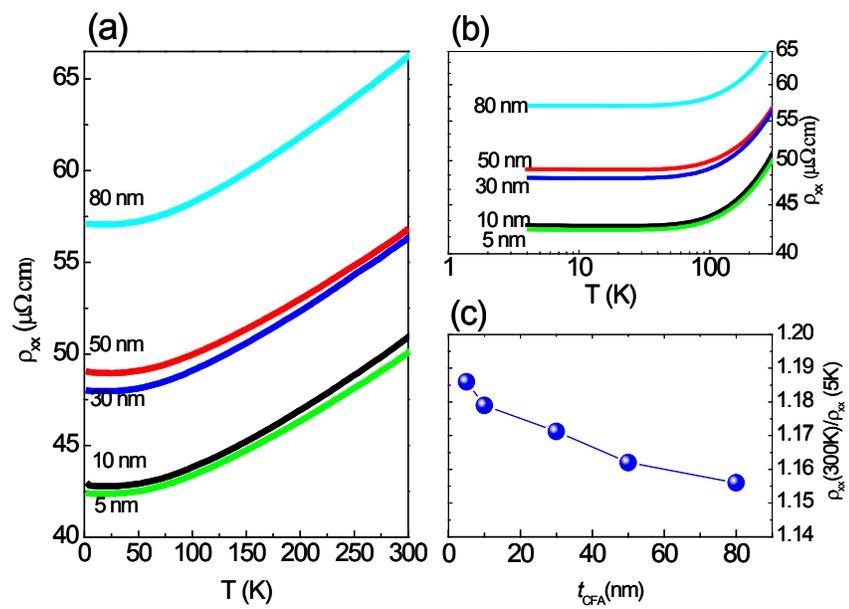

FIG. 4

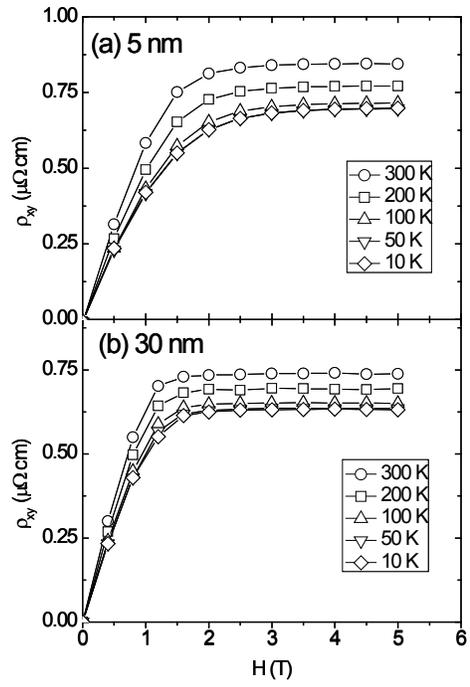

FIG. 5

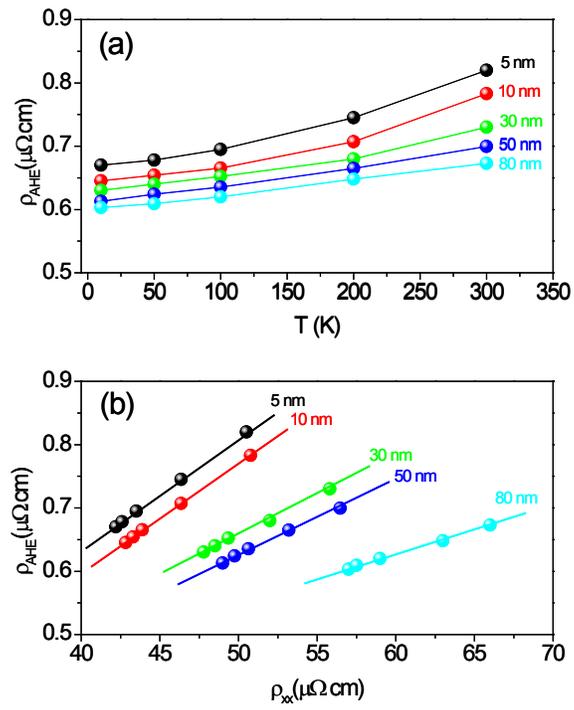

FIG. 6